\documentclass[preprint,aps,showpacs,prd,tightenlines,twocolumn,10pt]{revtex4}

\usepackage{graphicx}
\usepackage[latin1]{inputenc}
\usepackage{bbm}
\usepackage{amsmath}

\newcommand{\be}{\begin{eqnarray}} 
\newcommand{\ee}{\end{eqnarray}}
\newcommand{\nn}{~\nonumber \\}

\newcommand{\ssh}{\hskip 0.6mm\not\hskip -0.6mm}
\newcommand{\SSH}{\hskip 0.9mm\not\hskip -0.9mm}

\newcommand{\bmp}{\noindent\begin{minipage}{16cm}}
\newcommand{\emp}{\end{minipage}\vskip 7mm} 

\newcommand{\adj}{\check}

\newcommand{\titel}[1]{}

\begin{document}


\title{Non-abelian vector backgrounds with restored Lorentz invariance}

\author{Dennis D. Dietrich}
\affiliation{The Niels Bohr Institute, Copenhagen, Denmark}

\date{July 9, 2005}


\begin{abstract}

The influence of vector backgrounds with restored Lorentz invariance on
non-abelian gauge field theories is studied. Lorentz invariance is ensured by 
taking the average over a Lorentz invariant ensemble of background vectors. 
Like in the abelian case \cite{dh}, the propagation of fermions is suppressed 
over long distances. Contrary to the fermionic sector, pure gauge 
configurations of the background suppress the long-distance propagation of 
the bosons only partially, i.e.~not beyond the leading contribution for a 
large number of colours.


\end{abstract}



\maketitle


\section{Introduction}

Apart from appearances and applications in many other fields of physics, 
vector backgrounds can be used in order to incorporate mass dimension two 
vector condensates into gauge field theories in a Lorentz invariant way
\cite{dh,hp,ls}
or, in a different context, in theories breaking Lorentz invariance explicitely 
\cite{lv}.
The latter approach is motivated from string \cite{string} and
non-commutative field theory \cite{noncom}.
Of late, various aspects of mass dimension two condensates have attracted 
attention (see e.g.~\cite{a2}). In order to incorporate the condensates'
influence, the gauge field is shifted by a background vector $\Phi$. In
general it is taken to be constant \cite{dh,hp,ls}, thereby corresponding to 
the long wave-length limit of a more general approach.
A homogeneous background alone does not break translational invariance. In 
case translational invariance should be broken one would have to average over 
all translations in order to restore it. Be this as it may, in any single background 
Lorentz invariance is violated. It has to be reinstated by taking the mean over 
a Lorentz invariant ensemble of background vectors. A Lorentz invariant ensemble is 
a set of vectors which is mapped onto itself under any Lorentz transformation,
while, of course, almost every single element changes.
In \cite{dh,hp,ls} it has been demonstrated that the elementary fermions
and/or bosons are removed from the spectrum of asymptotically freely 
propagating particles. Usually the ensembles of backgrounds used in the
literature \cite{hp,ls} contain gauge field
configurations leading to zero as well as to non-zero field tensors.
However, the work presented in \cite{dh} shows that the propagation of
fermions is already stopped in pure gauge configurations of the background
in Euclidean as well as in Minkowski space.
The present article goes beyond the abelian limit and investigates the
influence of non-abelian pure gauge backgrounds on the propagation of
quarks and gluons. Here also the restoration of gauge invariance is an
issue.

Section \ref{Weight} discusses the Lorentz invariant weight functions
characterising the ensembles of background vectors for non-abelian gauge
groups and the preservation of gauge invariance.
Section \ref{QCD} investigates the influence of the backgrounds on quarks
and gluons for a \mbox{$SU(3)$} gauge group. The main ingredients of the
theory, i.e.~the generating functional for the Green functions (section
\ref{GF}) and the two-point functions (section \ref{2PF}) are treated. 
Section \ref{Final} summarises the results.


\section{Weight classification\label{Weight}}

At the beginning let us remember some facts about the abelian case \cite{dh}
also needed for the non-abelian theory.
The Lorentz invariant ensemble of backgrounds is 
characterised by a weight function \mbox{$W(\Phi)$} which appears in the 
averaging prescription (\mbox{$\int_\Phi:=\int d^4\Phi$}):
\be
\langle{\cal O}\rangle_W
=
\int_\Phi W(\Phi) {\cal O},
\ee
which does not change under Lorentz transformations and is normalised in such 
a way that:
\be
\int_\Phi W(\Phi)=1.
\label{normalisation}
\ee
Except for when \mbox{$\Phi=0$}, which leads to the unmodified theory, all 
other Lorentz invariant quantities depending on the vector $\Phi$ must be 
functions of $\Phi^2$. Thus every allowed weight function $W$ can be cast into 
the form:
\be
W(\Phi)=c\delta^{(4)}(\Phi^2)+w(\Phi^2),
\ee
where $w$ is a normalisable function of $\Phi^2$.

In euclidean space \mbox{$\Phi^2=0$} also means \mbox{$\Phi=0$}, whereby that 
case could be encoded within $w_{\mathrm E}(\Phi^2)$, where the subscript 
$_\mathrm E$ marks the euclidean space. However, here, the possible 
contribution from \mbox{$\Phi=0$}, from the unmodified theory is to remain
marked clearly and the delta term is kept explicitely. Thereby the 
normalisation condition (\ref{normalisation}) becomes:
\be
\pi^2\int_0^\infty v~dv~w_{\mathrm E}(v)=1-c,
\ee
with \mbox{$v:=\Phi^2$}. The Lorentz invariant weight functions can be 
decomposed into elementary delta weights
\be
w^{\mathrm E}_{\lambda}(\Phi^2)
:=
(4\pi\lambda)^{-1}\delta(\Phi^2-\lambda),
\label{elementaryeuclid}
\ee
according to:
\be
w_{\mathrm E}(\Phi^2)
=
\int4\pi\lambda~d\lambda~
w^{\mathrm E}_{\lambda}(\Phi^2)~
w_{\mathrm E}(\lambda).
\ee

If in Minkowski space a time ordered formalism is to be pursued the basis
should be chosen differently. From
\be
2\pi\mathrm i\delta(\Phi^2-\lambda)=S_\lambda^-(\Phi)-S_\lambda^+(\Phi),
\ee
with the time ordered ($+$) and anti time-ordered ($-$) scalar propagators
\be
S_\lambda^\pm(\Phi)=(\Phi^2-\lambda\pm i\epsilon)^{-1},
\ee
follows that the elementary weight function of choice is constructed from the
scalar Feynman propagator \mbox{$S^+_\lambda(\Phi)$}. Yet in Minkowski space
the hyperboloid pair defined by \mbox{$\Phi^2=\mathrm{const.}$} has infinite
content. Thus the minimal normalisable weight is given by the sum of three 
terms \cite{dh}:
\be
w_{\mathrm M}(\Phi^2)=\sum_{j=1}^{3}a_j S^+_{\lambda_j}(\Phi),
\label{elementaryminkowski}
\ee
satisfying:
\mbox{$
\sum_{j=1}^3a_j=0
$},
\mbox{$
\sum_{j=1}^3a_j\lambda_j=0,
$}
and the normalisation condition:
\mbox{$
(4\pi^2/4)\sum_{j=0}^3a_j\lambda_j\ln{\lambda_j}=1-c
$}.
The subscript $_M$ stands for "Minkowski".
On top of that, the case \mbox{$\Phi=0$} cannot be described by a function 
\mbox{$w_{\mathrm M}=w_{\mathrm M}(\Phi^2)$} and must be added in form of a 
delta term where required.


\subsubsection{Non-abelian}

In the present article, already at this point the background 
vector is identified with a component of the gauge field and the approach is 
generalised to pure gauge configurations in non-abelian gauge groups.
In the non-abelian case the gauge field carries colour. Nevertheless,
the different Lorentz components of a homogeneous pure gauge background 
commute. The background is to transform homogeneously under gauge 
transformations $U$ \cite{hp}:
\mbox{$
\Phi^a\rightarrow U^\dagger\Phi^aU.
$}
Therefore one can decompose the background (for the fundamental 
representation) according to:
\be
\Phi_\mu=\sum_{a=1}^{N_{\mathrm c}}U^\dagger\Theta^aU\Phi_\mu^a,
\ee
where the $\Theta^a$ are $N_{\mathrm c}$ projectors. Only 
summations on colour indices are carried out, which are marked explicitely.
The projectors satisfy 
\mbox{$\Theta^a\Theta^b=\delta^{ab}\Theta^a$}, where $\delta^{ab}$ is the 
Kronecker symbol. Thus a function of the background vector $\Phi$ can be 
deconstructed in the basis of the projectors $\Theta^a$:
\be
f(\Phi)=\sum_{a=1}^{N_{\mathrm c}}U^\dagger\Theta^aUf(\Phi^a).
\ee
The gauge dependence resides entirely in the transformations $U$ of the
projectors $\Theta^a$. The $\Phi^a$ being gauge invariant quantities one can
study each addend separately. Averaging each of them over its proper
weight yields:
\be
\langle f(\Phi)\rangle_{\{\Lambda^a\}}
=
\sum_{a=1}^{N_{\mathrm c}}\Theta^a_U
\int_{\Phi^a}
W_{\Lambda^a}(\Phi^a)
f(\Phi^a),
\label{average}
\ee
where
\mbox{$\Theta^a_U:=U^\dagger\Theta^aU$}. This is the same as introducing the
product weight function:
\be
W_{\{\Lambda^a\}}(\Phi)
:=
\prod_{a=1}^{N_{\mathrm c}}
W_{\Lambda^a}(\Phi^a)
\ee
and integrating over the $(\mathbbm{R}^4)^{N_{\mathrm c}}$, because every
single weight $W_{\Lambda^a}(\Phi^a)$ is already normalised.

In situations where the gauge invariance of the resulting quantity is required,
it can be ensured by averaging over all gauge transformations $U$ of the 
background $\Phi$:
\be
u\langle U^\dagger\Theta^a U\rangle_U
:=
\int [dU]U^\dagger\Theta^a U
=
uN_{\mathrm c}^{-1},
\ee
where $u$ stands for the volume of the gauge group. The result is proportional 
to unity in the corresponding space, because all non-singlet contributions
drop out. That result finally leads to:
\be
\langle\langle f(\Phi)\rangle_{\{\Lambda^a\}}\rangle_U
=
N_{\mathrm c}^{-1}\sum_{a=1}^{N_{\mathrm c}}
\int_{\Phi^a}
W_{\Lambda^a}(\Phi^a)
f(\Phi^a).
\label{averageu}
\ee
If all $\Lambda^a$ are the same the abelian case is recovered. Gauge
invariance can already be achieved by taking the average with the weight
$W$ if it, in itself, is gauge invariant. In the event where $W$ is
gauge invariant calculating the mean over the gauge group does not change
the result anymore.

For the fundamental representation of a \mbox{$U(N_{\mathrm c})$} gauge group 
all ingredients for the averaging procedure have been presented above. In the 
case of a \mbox{$SU(N_{\mathrm c})$} gauge group an additional constraint 
arises from the tracelessness of the generators:
\mbox{$
\sum_{a=1}^{N_{\mathrm c}}\Phi^a=0.
$}
Its incorporation leads to a coupling of the channels and thereby to a
modification of Eq.~(\ref{average}) which can be expressed 
by the weight function:
\be
W_{\{\Lambda^a\}}^{SU}(\Phi)
:=
\delta^{(4)}\left(\sum_{a=1}^{N_{\mathrm c}}\Phi^a\right)
\times
\prod_{a=1}^{N_{\mathrm c}-1}
W_{\Lambda^a}(\Phi^a).
\label{averagesu}
\ee


\subsubsection{Adjoint representation}

The background for bosonic correlators transforms under the adjoint
representation of the gauge group. The adjoint representation of the
\mbox{$SU(N_{\mathrm c})$} can be embedded in the fundamental of the 
\mbox{$SU({N_{\mathrm c}}^2-1)$} or the \mbox{$U({N_{\mathrm c}}^2-1)$}.
At the end the relevant result is extracted by imposing additional
constraints. In general, members of the adjoint representation of
$SU(N_{\mathrm c})$ are
hermitian and antisymmetric. Therefore the eigenvalues either vanish or come
in pairs with opposite sign \cite{gcn}. 

For the adjoint representation of \mbox{$SU(2)$} embedded in \mbox{$U(3)$} 
this means that the combination of eigenvalues \mbox{$\phi^1=-\phi^2$} and 
\mbox{$\phi^3=0$} always exists simultaneously. 
For the adjoint representation of \mbox{$SU(3)$} embedded in \mbox{$U(8)$} 
one has
\mbox{$\phi^1=-\phi^2$}, \mbox{$\phi^3=-\phi^4$}, \mbox{$\phi^5=-\phi^6$}, 
and \mbox{$\phi^7=0=\phi^8$} \cite{gcn}. 
This induces the following form for the average:
\be
\langle f(\phi)\rangle_{\{\lambda^a\}}^{SU_3^{{\mathrm adj.}}}
&=&
\sum_{a\in\cal M}
\int_{\phi^a}
W_{\lambda^a}(\phi^a)
(\theta^a_U+\theta^{a+1}_U)f(\phi^a)
+
\nn
&&+
(\theta^7_U+\theta^8_U)f(0),
\label{averageadj}
\ee
with $\theta^a$ the $U({N_{\mathrm c}}^2-1)$ projectors, 
\mbox{${\cal M}:=\{1;3;5\}$}, and where use has been made of the fact that 
$W$ is an even function of $\phi^a$.
In the case of the adjoint representation taking the average over the gauge
group yields a prefactor \mbox{$({N_{\mathrm c}}^2-1)^{-1}$} instead of
${N_{\mathrm c}}^{-1}$.


\section{QCD with the background\label{QCD}}

\subsection{Generating functional\label{GF}}

The generating functional for the time-ordered Green functions of QCD reads:
\be
Z=Z_\mathrm{int}Z_A Z_\chi Z_\psi 
\ee
with the interaction part \mbox{$Z_\mathrm{int}$}, whose non-fermionic part
depends on the chosen gauge and will not be specified here. The other
(free) factors are the bosonic
\be
Z_A=
\exp\left\{
\frac{i}{2}\int_{x,y}
J(x)\cdot\Gamma_0(x-y)\cdot J(y)
\right\},
\ee
the ghost
\be
Z_\chi=
\exp\left\{
-i\int_{x,y}
\xi^*(x)\Gamma_0(x-y)\xi(y)
\right\},
\ee
and the fermionic part
\be
Z_\psi=
\exp\left\{
-i\int_{x,y}
\bar{\eta}(x) G_0(x-y)\eta(y)
\right\},
\ee
with the free boson propagator \mbox{$\Gamma_0^{\mu\nu}(x-y)$}, the free
ghost propagator \mbox{$\Gamma_0(x-y)$}, the free fermion propagator
\mbox{$G_0(x-y)$}, as well as the currents $J$, $\xi$, $\xi^*$, $\eta$, and
$\bar{\eta}$.

The background is included in the theory by translating the gauge field $A$
by the background $\Phi$---\mbox{$A_\mu\rightarrow A_\mu+\Phi_\mu$}---and 
afterwards taking the mean over the ensemble of these vectors and restoring
gauge invariance with respect to gauge transformations of the background. 
Doing so leads to the modified generating functional:
\be
{\cal Z}
=
\langle\langle 
Z_{\mathrm{int}}^\Phi Z_A^\Phi Z_\chi^\Phi Z_\psi^\Phi
\rangle_W\rangle_U,
\label{mean}
\ee
with the different parts evaluated in a single background $\Phi$. The
non-interacting factors are obtained by replacing the free two-point
functions by those in a single background $\Phi$. They shall be defined in
the next section.
\mbox{$Z_{\mathrm{int}}^\Phi$} contains all terms of
third and fourth order in the dynamic fields. Its $\Phi$ dependence
originates from the four-gluon vertex with three dynamical gluon legs and
one coupling to the background.
The propagators studied in the following section are obtained by taking the
functional derivatives of the above generating functional with respect to
the corresponding currents.

In Eq.~(\ref{mean}) a common mean over the sectors of the generating 
functional transforming under the fundamental and the adjoint representation 
of the gauge group, respectively, is taken. For this reason a connection has
to be made between the averages calculated with the weights (\ref{averagesu}) and
Eq.~(\ref{averageadj}), respectively. Especially this fact is important if
correlators are to be calculated that involve bosonic and fermionic fields
simultaneously. In $SU(3)$ one has the connection \cite{gcn}:
\be
2(\phi^1_\mu)^2
&=&
{\Phi_\mu}^2
[1-\cos(\theta_\mu)],
\nn
2(\phi^3_\mu)^2
&=&
{\Phi_\mu}^2
[1+\cos(\theta_\mu-\pi/3)],
\nn
2(\phi^5_\mu)^2
&=&
{\Phi_\mu}^2
[1+\cos(\theta_\mu+\pi/3)],
\ee
with
\be
1+\cos^3(\theta_\mu)=18({\Phi_\mu}^2)^{-3}\prod_{a=1}^3(\Phi^a_\mu)^2.
\ee
This relationship holds separately for each Lorentz component, which is why 
one does not sum over
the index $\mu$ anywhere in the previous equations. These relations allow to
construct the corresponding weight for one representation from the weight 
function for the other.


\subsection{Two-point functions\label{2PF}}

A principal ingredient of the modified generating functional and thereby of
the modified theory are the two-point correlators in the background. They 
also serve to establish the link to the results presented in \cite{dh}.

With the usually made assumption that all other condensates be absent
\cite{hp,ls} the fermionic propagator in the background obeys the equation of 
motion
\be
[i\ssh\partial(x)+\SSH\Phi-m]G_\Phi(x-y)=\delta^{(4)}(x-y).
\ee
Its solution in a pure gauge background is given by:
\be
G_\Phi(z)=e^{i\Phi\cdot z}G_0(z),
\label{fermionphi}
\ee
where \mbox{$G_0(z)$} is the solution of the free equation.

The bosonic and ghost propagators depend on the chosen gauge.
The gauge-fixing contribution to the lagrangian density in background-field 
Feynman gauge is given by (this time summing over all colour indices):
\be
{\cal L}_{\mathrm GF}
=
-
\frac{1}{2}
\{[\partial_\mu-i\Phi_\mu]^{ab}A^{b\mu}\}
\{[\partial_\nu-i\Phi_\nu]^{ac}A^{c\nu}\},
\ee
which is Lorentz covariant. Without further spontaneous 
symmetry breaking, i.e.~with \mbox{$\langle A\rangle=0$}, the equation of 
motion for the gluon propagator reads \cite{glue}:
\be
[\partial(x)-i\adj\Phi]\cdot[\partial(x)-i\adj\Phi]
\Gamma^{\mu\nu}_\Phi(x-y)
=
\delta^{(4)}(x-y)g^{\mu\nu}
\ee
where matrix multiplication of the colour matrices is understood implicitely
and the $\adj~$ indicates that the adjoint representation has to be used. In
the same gauge the ghost propagator obeys the equation of motion \cite{glue}:
\be
[\partial(x)-i\adj\Phi]\cdot[\partial(x)-i\adj\Phi]\Gamma_\Phi(x-y)
=
\delta^{(4)}(x-y).
\ee
Therefore one has in background field Feynman gauge:
\be
\Gamma_\Phi^{\mu\nu}(x-y)=g^{\mu\nu}\Gamma_\Phi(x-y)
\ee
and it is sufficient to study one of the two propagators, e.g.~the ghost
propagator \mbox{$\Gamma(x-y)$}. In a pure gauge background one has then:
\be
\Gamma_\Phi(z)=e^{i\adj\Phi\cdot z}\Gamma_0(z),
\label{ghostphi}
\ee
with \mbox{$\Gamma_0(z)$} the solution of the background-free equation.

As had already been seen in the abelian case \cite{dh} the Fourier phases of
the propagators in the background $\Phi$ lead to the averaging procedure
being identical to a Fourier transformation of the weight function. For this
reason, in coordinate space the $n$-point functions of the modified theory
are the $n$-point functions of the theory without background
multiplied by a $n$-point function which is the Fourier transformed weight.
The latter is a genuine $n$-point function. That means that even if in the
original theory the higher correlators factorise into lower ones, they 
will not
in the modified theory. Thereby especially if the original theory should
display gaussianity on a certain level, the modified theory will not.

There are other situations where such field configurations play a r\^ole.
The zero components of aforesaid Fourier phases resemble chemical
potentials \cite{dh}. The spatial components are similar to what one
encounters for twisted boundary conditions on compact spaces \cite{b}.
In a colour superconductor a field configuration corresponding to a zero 
field tensor serves to restore its colour neutrality \cite{dr}.

The theory in the pure gauge background can be seen as one with a modified 
vacuum structure without background energy density. One could express this
by writing:
\be
\left.\left<\langle 0|
T\psi^n\bar{\psi}^nA^{n^\prime}
|0\rangle\right|_\Phi\right>_W
=:
\langle\Omega|
T\psi^n\bar{\psi}^nA^{n^\prime}
|\Omega\rangle
,
\ee
where \mbox{$|\Omega\rangle$} represents the non-trivial vacuum. The vacuum
expectation values of the modified theory
\mbox{$\langle\Omega|{\cal O}|\Omega\rangle$} are the averaged vacuum
expectation values 
\mbox{$\langle\langle 0|{\cal O}|0\rangle|_\Phi\rangle_W$} each in a 
single background.


\subsubsection{Euclidean space}

Computing the average of the fermionic propagator (\ref{fermionphi}) according to 
Eq.~(\ref{averageu}) \mbox{[$U(N_{\mathrm c})$]} with the elementary weights
(\ref{elementaryeuclid}), choosing \mbox{$c=0$}, and averaging over the gauge group
leads to:
\be
\langle
\langle G_\Phi(z)\rangle_{\{\Lambda^a\}}^{\mathrm E}
\rangle_{\mathrm U}
=
N_{\mathrm c}^{-1}
\sum_{a=1}^{N_{\mathrm c}}
\frac{\sin\sqrt{\Lambda^a z^2}}{\sqrt{\Lambda^a z^2}}
G_0(z)
\label{fermionelementary}
\ee
In a $SU(N_{\mathrm c})$ gauge group the result reads [see
Eq.~(\ref{averagesu})]:
\be
\langle
\langle G_\Phi(z)\rangle_{\{\Lambda^a\}}^{\mathrm E, SU}
\rangle_{\mathrm U}
=
N_{\mathrm c}^{-1}
\sum_{a=1}^{N_{\mathrm c}-1}
\frac{\sin\sqrt{\Lambda^a z^2}}{\sqrt{\Lambda^a z^2}}
G_0(z)
+
\nn
+
{N_{\mathrm c}}^{-1}
G_0(z)
\prod_{a=1}^{N_{\mathrm c}-1}
\frac{\sin\sqrt{\Lambda^a z^2}}{\sqrt{\Lambda^a z^2}}
\ee
For large distances $\sqrt{z^2}$, the propagator is damped with respect
to the free one. At short distances $\sqrt{z^2}$ the free propagator
is recovered. In momentum space this manifests itself in the
on-shell pole being removed and a pole proportional to \mbox{$1/\sqrt{k^2}$}
being introduced \cite{dh}. The elementary fermions are no longer part of the
spectrum of freely propagating particles.

This result resembles the one for the tree-level propagator in \cite{hp,ls}. 
However, there, as opposed to here configurations of the gauge field were 
taken into account that lead to a non-vanishing field tensor. The weight
used there \mbox{$W_\mathrm{HP}(\Phi)\sim\exp(-\Phi^2/\Lambda^2)$} but 
constrained to zero field tensors in coordinate space gives
\mbox{[$U(N_{\mathrm c})$]}:
\be
\langle\langle G_\Phi(z)\rangle_\mathrm{HP}^\mathrm{E}\rangle_U
=
\exp[-z^2\Lambda^2/4]G_0(z),
\ee
or in a \mbox{$SU(N_{\mathrm c})$} gauge group:
\be
\langle\langle G_\Phi(z)\rangle_\mathrm{HP}^{\mathrm E, SU}\rangle_U
=
\{(1-{N_{\mathrm c}}^{-1})
\exp[-z^2\Lambda^2/4]
+
\nn
+
{N_{\mathrm c}}^{-1}
\exp\left[-z^2\Lambda^2(N_{\mathrm c}-1)/4\right]\}G_0(z),
\ee
which for a large number of colours $N_{\mathrm c}$ reduces to the 
\mbox{$U(N_{\mathrm c})$} result.
Again the propagation over large distances $\sqrt{z^2}$ is suppressed while
the free propagator is obtained for \mbox{$\sqrt{z^2}\rightarrow 0$}.

As Eq.~(\ref{fermionphi}) is satisfied by every fermion propagator which is a
singlet of the colour group, the previous result holds for those
correlators, too. If a fermionic two-point function \mbox{$g(x-y)$}
should not be colour neutral, in general, the projectors $\Theta^a$ do not 
commute with the correlator. This leads to a breaking of translational 
invariance:
\be
g_\Phi(x-y)
=
e^{+i\Phi\cdot x}g(x-y)e^{-i\Phi\cdot y}
=
\nn
=
\sum_{a,b=1}^{N_{\mathrm c}}
e^{i(\Phi^a\cdot x-\Phi^b\cdot y)}\Theta^ag(x-y)\Theta^b
\ee
As mentioned above it usually would have to be restored by averaging over all 
translations.
However here it is already required to take the average over the gauge group 
anyhow. This
is equivalent---up to a factor---to taking the trace over the colour matrices:
\be
N_{\mathrm c}\langle\Theta^ag(z)\Theta^b\rangle_U
=
\mathrm{tr}\{\Theta^ag(z)\Theta^b\}
=
\mathrm{tr}\{\Theta^ag(z)\}\delta^{ab}
\ee
which is non-zero only for \mbox{$a=b$} due to the projection properties of
the $\Theta^a$. Therefore
translational invariance is restored. Enforcing Lorentz invariance, for
example, in the case of a \mbox{$U(N_{\mathrm{c}})$} gauge group leads to 
Eq.~(\ref{fermionelementary}) but where the free propagators \mbox{$G_0(z)$} are 
replaced by \mbox{$\mathrm{tr}\{\Theta^ag(z)\}$}. That means that also here 
the propagation is stopped at large distances and almost free at short
distances.



In background field Feynman gauge for the present investigation it is
sufficient to study either the gluon or the ghost propagator. Let us take the
latter. Calculating the average of the ghost propagator (\ref{ghostphi}) 
according to Eq.~(\ref{averageadj}) with the
elementary weights (\ref{elementaryeuclid}), choosing \mbox{$c=0$}, and averaging 
over the gauge group yields:
\be
\langle\langle
\Gamma_\Phi(z)
\rangle_{\{\lambda^a\}}^{\mathrm E,SU_3^{\mathrm{adj.}}}\rangle_U
=
\frac{2}{8}\Gamma_0(z)
\left(
1+\sum_{a\in\cal M}
\frac{\sin\sqrt{\lambda^a z^2}}{\sqrt{\lambda^a z^2}}
\right).
\ee
The free propagator is recovered for short distances $\sqrt{z^2}$. Six of
eight channels are suppressed at large distances $\sqrt{z^2}$ while the two
belonging to the zero eigenvalues remain freely propagating. Contrary to the
fermions, which belong to the fundamental representation of the gauge group,
the ghosts and gluons can be kept from propagating only partially.
In momentum space the suppressed channels behave like the
fermionic propagators in the background, i.e.~the on-shell pole is absent
and a pole \mbox{$1/\sqrt{k^2}$} present.

These results hold for every ghost/gluon propagator that is a
colour singlet. If it is not, translational invariance is restored due to the
average over the gauge group being taken, with the analogous consequences as 
for the non-singlet propagator in the fermionic case. 

In \cite{hp} for gluons in background configurations with, in general, 
non-zero field tensor and evaluated at leading order in
\mbox{${N_\mathrm{c}}^{-1}$} no residue of freely propagating bosons is
found. Already for fundamental reasons this cannot be due to the specific 
form of the weight. This can also be seen directly because two free channels remain again in the 
spectrum, if it is used within the present framework:
\be
\langle\langle
\Gamma_\Phi(z)
\rangle_\mathrm{HP}^{\mathrm E,SU_3^\mathrm{adj.}}\rangle_U
=
\frac{2}{8}
\Gamma_0(z)
\left\{
1+3\exp[-z^2\lambda^2/4]
\right\}.
\ee
Thus the origin of this difference with respect to \cite{hp} could be the 
non-zero field tensors. 
In the case where the Lorentz components of the
background vector do not commute---i.e.~if the field strength is
finite---a simultaneous diagonalisation of all Lorentz components is
impossible. Then products of matrices can occur which do not belong to the
adjoint representation, which is not closed with respect to matrix
multiplication of its members. Thereby the necessarily existing zero
eigenvalues of the adjoint representation could be avoided and the propagation
of the gluons be stopped entirely. However as the number of eigenvalue channels in the adjoint representation 
equals \mbox{${N_\mathrm{c}}^2-1$} and the multiplicity of zero 
eigenvalues is \mbox{$N_\mathrm{c}-1$} they are of subleading
importance for large values of $N_\mathrm{c}$ and go unnoticed in an
analysis based on the leading order of an expansion in ${N_\mathrm{c}}^{-1}$.


\subsubsection{Minkowski space}

Taking the average of the fermionic propagator (\ref{fermionphi}) according to 
Eq.~(\ref{averageu}) \mbox{[$U(N_{\mathrm c})$]} with the elementary weights
(\ref{elementaryminkowski}), choosing \mbox{$c=0$}, and averaging over the gauge group
yields:
\be
\langle\langle
G_\Phi(z)
\rangle_{\{\Lambda^a_j\}}^\mathrm{M}\rangle_U
=
{N_\mathrm{c}}^{-1}
\sum_{a=1}^{N_\mathrm{c}}
\sum_{j=1}^3
a^a_j
s^+_{\Lambda^a_j}(z)
G_0(z),
\ee
with 
\mbox{$
s^+_\Lambda(z)
:=
4\pi^2\sqrt{\Lambda}
{\mathrm{K}_1(\sqrt{\Lambda}\sqrt{-z^2+i\epsilon})}
/
{\sqrt{-z^2+i\epsilon}}
$}.
In a $SU(N_{\mathrm c})$ gauge group the result becomes [see
Eq.~(\ref{averagesu})]:
\be
\langle\langle
G_\Phi(z)
\rangle_{\{\Lambda^a_j\}}^{\mathrm{M},SU}\rangle_U
=
{N_\mathrm{c}}^{-1}
\sum_{a=1}^{N_\mathrm{c}-1}
\sum_{j=1}^3
a^a_j
s_{\Lambda^a_j}^+(z)
G_0(z)
+
\nn
+
{N_\mathrm{c}}^{-1}
\prod_{a=1}^{N_\mathrm{c}-1}
\left[
\sum_{j=1}^3
a^a_j
s_{\Lambda^a_j}^+(z)
\right]
G_0(z).
\ee
For \mbox{$z^2=0$} the previous expressions reduce to the free propagator.
For large absolute values of $z^2$ the propagator is suppressed at least
proportionally to \mbox{$|z^2|^{-3/4}$} \cite{dh} with respect to the free
propagator. Therefore the fermions cannot propagate over arbitrarily large
distances.

Like in the abelian case all terms for the
\mbox{$U(N_\mathrm{c})$} gauge group correspond to contributions of scalars 
to the self energy of the fermion without external legs, thereby indicating
that no freely propagating particles are described. For the 
\mbox{$SU(N_\mathrm{c})$} gauge group the first terms have the same
interpretation. Only the last one corresponds to a sum over multiple
\mbox{$(N_\mathrm{c}-1)$}-loop contributions of scalars to the self energy of
the fermion once more without external legs. Thus what was said concerning the 
absence of freely propagating particles remains valid.

Computing the average of the ghost propagator (\ref{ghostphi}) 
according to Eq.~(\ref{averageadj}) with the
elementary weights (\ref{elementaryminkowski}), choosing \mbox{$c=0$}, and averaging 
over the gauge group yields:
\be
\langle\langle
\Gamma_\Phi(z)
\rangle_{\{\lambda^a_j\}}^{\mathrm{M},SU_3^{\mathrm{adj.}}}\rangle_U
=
\frac{\Gamma_0(z)}{4}
\left[
1
+
\sum_{a\in\cal M}
\sum_{j=1}^3
a^a_j
s_{\lambda^a_j}^+(z)
\right].
\ee
For small absolute values of $z^2$ the free propagator is reobtained. At
large absolute values of $z^2$ two \mbox{(${N_\mathrm{c}}-1$)} of the eight
\mbox{(${N_\mathrm{c}}^2-1$)} channels remain unaltered
and correspond to freely propagating gluons.

The terms for which the propagation over long
distances is suppressed again do resemble the contribution of scalars to the
self energy of the considered particle---which now is the gluon/ghost---without
external legs. 

The observations for ghosts and gluons remain the same in all background
field Lorenz gauges. For those the ghost propagator (\ref{ghostphi}) stays 
the same. In the gluon propagator only the free part 
\mbox{$\Gamma^{\mu\nu}_0(z)$} 
changes, which leaves the envelope function \mbox{$\widetilde{W}(z)$} unchanged. 

Also in Minkowski space the above findings are valid for every propagator 
that is a colour singlet. For non-singlet propagators, translational 
invariance is 
recovered after the average over the gauge group has been taken. The
consequences remain the same as in euclidean space.


\section{Summary\label{Final}}

Non-abelian gauge field theories have been studied whose correlators are defined with an
additional average over a Lorentz invariant ensemble of homogeneous vector
backgrounds. The ensembles are constrained to pure gauge configurations of
the background. For the fermions their presence suffices in order to
ascertain that they do not propagate freely over large distances at all,
neither in euclidean nor in Minkowski space.
This finding coincides with the one for abelian gauge groups \cite{dh} and
the one obtained admitting non-zero field tensors \cite{hp,ls}. For the
gluons the propagation over long distances can only be stopped in some
channels. Those remaining free have zero eigenvalues which are always
present in the adjoint representation but subleading in
${N_\mathrm{c}}$. The gauge fields as
opposed to the fermions have to transform under the adjoint representation as 
a consequence of the requirement of gauge invariance, whence the presence of 
channels of propagation escaping suppression over long distances in this
framework is unavoidable for the gauge bosons.
Naturally, in the abelian case the bosonic propagator was unaltered altogether 
\cite{dh}. The results in the present article
for the propagators in pure gauge backgrounds do not only hold for the
correlators to all orders in the background and otherwise at tree-level but
for every colour singlet or non-singlet propagator and consequently also for 
the full one, i.e.~to all loops.

In Minkowski space the contributions which do not describe freely propagating 
particles resemble contributions of scalars to the self energy of the
particle without external legs. This fact confirms the interpretation
concerning the non-propagation.

The feature that the propagation over short distances
remains essentially unchanged while it is modified over long distances is
shared by non-commutative field theories preserving Lorentz invariance 
\cite{nonbreak}.


\vskip 3mm

I would like to thank
Adrian Dumitru,
Ralf Hofmann, 
Stefan Hofmann, 
Andrew D.~Jackson,
Gregory Korchemsky, 
Amruta Mishra, 
Kerstin Paech,
Olivier P$\grave{\mathrm e}$ne, 
Joachim Reinhardt, 
Dirk Rischke, 
Francesco Sannino, 
Stefan Schramm,
Igor Shovkovy, 
and
Kim Splittorff 
for inspiring and informative discussions.
Again my thanks are due to
Adrian Dumitru 
and 
Stefan Hofmann
for reading the manuscript.


\end{document}